\documentclass[11pt,a4paper]{article}
\pdfoutput=1
\usepackage{jcappub}

\usepackage{amssymb,amsmath,amsthm}
\usepackage{braket}
\usepackage[T1]{fontenc}
\usepackage[latin1]{inputenc}
\usepackage[english]{babel}

\usepackage{float}
\usepackage{braket}
\usepackage{slashed}
\usepackage{commath}
\usepackage{diagbox}
\usepackage{color}
\usepackage{bm}
\usepackage{booktabs}
\usepackage{mathrsfs}
\usepackage{enumitem}
\usepackage{subfigure}

\renewcommand{\epsilon}{\varepsilon}
\newcommand{\mb}[0]{\mathbf}

\newcommand{\p}{\partial}

\newcommand{\Ai}{\mathrm{Ai}}
\newcommand{\Bi}{\mathrm{Bi}}

\begin{document}
\title{Reheating via gravitational particle production in kination epoch}

\author{Juho Lankinen,}
\author{Oskari Kerppo}
\author{and Iiro Vilja}

\affiliation{Turku Center for Quantum Physics, Department of Physics and Astronomy, University of Turku, Turku 20014, Finland}

\emailAdd{jumila@utu.fi}
\emailAdd{oeoker@utu.fi}
\emailAdd{vilja@utu.fi}

\abstract{
We provide a detailed study of reheating in the kination regime where particle content is created by gravitational production of massive scalars mutually interacting with a massless scalar field. The produced particles subsequently decay into massless particles eventually reheating the Universe.  We aim for a more precise picture using Boltzmann equations and decay rates obtained by methods of quantum field theory in curved spacetime. By numerical calculations it is found that after inflation the Universe ends up being dominated by ordinary matter for a while before the radiation dominated era. The reheating temperature itself is found to be in the $10^6-10^{12}\ \text{GeV}$ regime.

}

\maketitle

\section{Introduction}

In the inflationary scenario, the Universe underwent a rapid expansion phase in the very early stages of its existence. After this expansion, the almost standard procedure for creating the particle content of the Universe is through decay of the oscillations of the inflaton \citep{Starobinsky:1980,Linde:1982} or by collision of bubbles \citep{Guth:1981}. However, one ultimately needs only a mechanism for creating radiation which eventually drives the expansion of the Universe. Some time ago it was realized that gravitational particle creation during the change of metric from de Sitter to post-inflationary metric could be used as a possible mechanism to reheat the Universe \citep{Spokoiny:1993,Ford:1987}. Ford  considered a change of metric from de Sitter to either radiation or matter dominated era \citep{Ford:1987}, while Spokoiny considered a scenario where the Universe ends up in a deflationary period \citep{Spokoiny:1993}. This period, also known in literature as kination \citep{Joyce:1997,Joyce_Prokopec:1998}, can be realized when the kinetic energy of the inflaton dominates its potential energy. Hence, in the kination regime the Universe ends up in a state characterized by perfect fluid with a stiff equation of state; the Universe is dominated by stiff matter. An early stiff matter era was presumably first considered by Zel'dovich \citep{Zeldovich:1972} and indeed, the stiff equation of state in of itself exhibits several interesting cosmological properties worth of further investigation \citep{Barrow:1978,Chavanis:2015}.
Reheating in a kination epoch has received quite a bit of interest lately and numerous new studies have been done \citep{deHaro:2016a,Chun:2009,Pallis:2006,Salo:2017a,Salo:2017b,Hashiba:2019}. Most of these studies are concerned with quintessence models incorporating a period of kination and the particle production is obtained through a  change of metric from de Sitter to post-inflationary metric.  Particle creation by a sudden or smooth change of metric is not the only mechanism of gravitational particle creation able to reheat the Universe though: it is also possible to create the matter by the very expansion of spacetime itself \citep{Parker:1969,Parker:1971,Parker:1968}.

With particle creation as a starting point, and contrast to previous studies, we consider a reheating scenario where matter is created during the expanding stiff matter dominated era by gravitational particle production instead of matter being produced by the sudden change of the metric from de Sitter to post-inflationary metric.
The gravitationally created particles decay into massless particles (radiation) which ultimately reheat the Universe. Particles of the standard model, or some other model, are eventually produced from these relativistic particles.
 It is assumed that the Universe starts empty after inflation, so no particles have been produced by a change of metric. It is reasonable to assume that the energy density created at the transition is much smaller than one created during the expansion as inflation inflates out existing matter. 
We will describe particle decay using quantum field theory in curved spacetime instead of using Minkowskian field theory. This aspect has attained renewed interest recently \citep{Lankinen_Vilja:2018a,Lankinen_Vilja:2018b,Lankinen_Vilja:2017b,Lankinen_Malmi:2019,Boyanovsky}. Decay in curved spacetime is conceptually very difficult and even though the typical in-out formalism might be mathematically sound, there are physical arguments which suggest that another method to calculate decay rates must be introduced to obtain physically meaningful results \citep{Audretsch_Spangehl:1985,Audretsch_Ruger_Spangehl:1987,Audretsch:1986}. Therefore, we will use a physically motivated method known as the added-up method \citep{Audretsch_Spangehl:1985} to calculate the decay rate. Another major feature of our study is the use of Boltzmann equations. Although the Boltzmann equations are used in an integrated form, together with the curved space quantum field theoretic aspects, the model we introduce in this paper presents a novel approach to study reheating and provides a way to give a more precise picture of the procedure.

This paper is structured in the following way. In section \ref{sec:2} we introduce our model and describe the assumptions made. The procedure of obtaining the reheating temperature is the topic of section \ref{sec:3} and the numerical results are presented in section \ref{sec:4}. Finally, we discuss about the results in section \ref{sec:5}.

\section{The Model}\label{sec:2}
In this section we introduce the individual pieces needed to describe the reheating scenario and discuss about the assumptions that have been made.

\subsection{Preliminaries}

The post-inflationary Universe is described by a four dimensional spatially flat Robertson-Walker metric of the form
\begin{align}
	ds^2=dt^2-a(t)^2 d\mathbf{x}^2
	\end{align}
given in standard coordinate time $t$ with a dimensionless scale factor $a(t)$. The metric signature is $+---$ throughout. The metric can also be given in terms of conformal time $\eta$,  defined by the relation $dt=a(\eta)d\eta$, as
	\begin{align}
	ds^2=a(\eta)^2(d\eta^2- d\mathbf{x}^2).
	\end{align}
We will use both standard and conformal time as is appropriate at any given instance.
For the conformal scale factor, we choose $a(\eta)=b\eta^{n/2}$ with $b$ a positive constant controlling the expansion rate of the Universe. In standard coordinate time, this scales as $a(t)\propto t^{n/(2+n)}$. As the scale factor is dimensionless, the units of the parameter $b$ depend on $n$. This parameter appears in the Boltzmann equations only through particle creation rate in the stiff matter period which fixes the units of $b$ to be $\text{GeV}^{1/2}$.

 The matter content of the Universe is described by a perfect fluid characterized by a dimensionless parameter $\omega$ through the equation $\omega=p/\rho$, where $p$ is the pressure and $\rho$ is the energy density of the fluid. With the scale factor given in conformal time, the parameter $n$ is related to $\omega$ by the equation \citep{Lankinen_Vilja:2018b}
\begin{align}
\omega=\frac{1}{3}\Big( \frac{4}{n}-1\Big).
\end{align}
Specifically, the values $n=1,2,4$ describe a universe which is dominated by stiff matter, radiation and ordinary matter, respectively.

In this background, we consider a propagating massive real scalar field $\phi$ with mass $m$ interacting with a massless scalar field $\chi$. 
The Lagrangian is given by
	\begin{align}\label{eq:Lagrangian}
	\mathcal{L}=&\frac{\sqrt{-g}}{2}\big\{\p_\mu \phi \p^\mu\phi-m^2\phi^2-\xi R\phi^2+\p_\mu\chi\p^		\mu \chi-\frac{1}{6}R\chi^2\big\}+\mathcal{L}_I,
	\end{align}
where $\mathcal{L}_I$ is the interaction term, $g$ stands for the determinant of the metric and $R$ is the Ricci scalar. The coupling $\xi$ has two special values. The value $\xi=1/6$ is known as conformal coupling in four dimensions and $\xi=0$ is known as minimal coupling. For the interaction term, we choose
\begin{align}\label{eq:InteractionTerm}
	\mathcal{L}_I=-\sqrt{-g}\lambda\phi\chi^2,
\end{align}
where $\lambda\neq 0$ is the coupling constant. At this point it is good to make few comments about the Lagrangian given in \eqref{eq:Lagrangian}. Apart from the massive scalar being coupled to gravity with an arbitrary coupling constant $\xi$, the massless particles are assumed to be conformally coupled. The reason for this is twofold. Since massless conformally coupled particles are not created by the expansion of spacetime \citep{Parker:1969}, only the creation of massive particles is to be taken into account simplifying the model considerably. Second, the calculation of particle decay rates in curved spacetime is inherently difficult and a method which enables the calculation, known as added-up method, requires conformally coupled massless particles as decay products in order for the decay rate interpretation to make sense in curved spacetime \citep{Audretsch_Spangehl:1985}.

\subsection{Gravitationally produced particles}
One of the phenomena arising from quantum field theory in curved spacetime is the gravitational creation of particles as a result of spacetime expansion. Because of the importance of this phenomena in our study, we give a very brief overview of it and refer the reader to \citep{Birrell_Davies,Parker_Toms} and references therein for a more thorough treatment of the subject. As the spacetime expands, the positive initial field modes $u^{\rm{in}}_{\mathbf k}(x)$ evolve into a linear combination of the field modes in the asymptotic future,
\begin{align}
u^{\rm{in}}_{\mathbf k}(x)=\alpha_k u^{\rm{out}}_{\mathbf k}(x)+\beta_k u^{\rm{out}*}_{\mathbf{ -k}}(x),
\end{align}
where $k=|\mb k|$ and $\alpha_k$ and $\beta_k$ are known as Bogoliubov coefficients and the superscript $\mathrm{out}$ refers to the modes in the asymptotic future. The connection between the Bogoliubov coefficients and particle creation is given by the relation
\begin{align}\label{eq:BogoSquared}
	\braket{N_k}_{t\to\infty}=|\beta_k|^2,
\end{align}
which gives the expectation value of number of particles present at late times in mode $k$. The square of the absolute value of the Bogoliubov coefficient $\beta_k$ can thus be interpreted as the probability of particles present at late times in mode $k$ which have been created by the expansion of spacetime. Moreover, we note that the differential creation rate can be obtained by differentiating equation \eqref{eq:BogoSquared} with respect to time.

At present, we are interested in the situation where particles are gravitationally created during the kination, or stiff matter dominated era. This situation was considered in \citep{Lankinen_Vilja:2017a}, where the differential creation rate in conformal time for a massive scalar particle in mode $k$ was obtained in terms of Airy functions as
	\begin{align}\label{eq:DiffCreation}
	\frac{d|\beta_k|^2}{d\eta}= &\frac{\pi z^2\eta}{2k}\Big[\Ai\Big( \frac{-k^2-z^2\eta}{z^{4/3}}\Big)\Ai'\Big( \frac{-k^2-z^2\eta}{z^{4/3}}\Big)+\Bi\Big( \frac{-k^2-z^2\eta}{z^{4/3}}\Big)\Bi'\Big( \frac{-k^2-z^2\eta}{z^{4/3}}\Big) \Big].
	\end{align}
In the above equation a shorthand notation $z=mb$ is used. This formula gives the differential particle creation rate which can be used in the Boltzmann equations. Next, we address the question of particle decay in curved spacetime.

\subsection{Decay in curved spacetime}
Decay rate calculations in curved space contain a lot of conceptual issues and technical difficulties, which may be one of the reasons Minkowskian decay rates are commonly used in cosmological calculations \citep{Kolb_Turner}. When curvature cannot be neglected anymore, Minkowskian decay rates are nevertheless only an approximation and curved space counterparts should be used instead. This problem has attained a wealth of interest in recent times and decay rates have been obtained for various processes \citep{Lankinen_Vilja:2017b,Lankinen_Vilja:2018a,Lankinen_Vilja:2018b}. 
For more on the conceptual and technical issues regarding particle decay in curved spacetime, we refer the reader articles \citep{Audretsch_Spangehl:1985,Lankinen_Vilja:2017b,Lankinen_Vilja:2018a,Lankinen_Vilja:2018b,Audretsch:1986,Audretsch_Ruger_Spangehl:1987,Audretsch_Spangehl:1986,Audretsch_Spangehl:1987}. For our purposes here it is enough to recall the results obtained in these articles which are necessary for our treatment of reheating, i.e., particle decay rates for stiff matter, radiation and matter dominated universes for the interaction \ref{eq:InteractionTerm}. In \citep{Lankinen_Vilja:2018b} the differential decay rate for a general power-law scale factor $a(\eta)=b\eta^{n/2}$ was obtained as
	\begin{align}\label{eq:DifferentialDecayRate}
	\Gamma_{\chi}(t)=\frac{\lambda^2t}{32} H^{(1)}_\alpha(mt)H^{(2)}_\alpha(mt),
	\end{align}
where $H^{(1,2)}_\alpha$ are the Hankel functions and the differential decay rate has been given in terms of the  standard coordinate time $t$. The index $\alpha$ is given as
	\begin{align}\label{alpha}
	\alpha:=\frac{\sqrt{1-n(n-2)(6\xi-1)}}{2+n}
	\end{align}	
and contains the index $n$ which gives the appropriate matter content of the Universe. It should be noted that equation \eqref{eq:DifferentialDecayRate} was derived for the special case in rest frame $\mb k=0$ of the massive particle. This was done in \citep{Lankinen_Vilja:2018b} in order to obtain an exact result for the decay rate. Regarding the use of equation \eqref{eq:DifferentialDecayRate} in the Boltzmann equations, it was shown in \citep{Lankinen_Vilja:2017a} that in a Universe dominated by stiff matter, particle production is peaked at low momentum and high mass modes. 
 Hence, the use of zero-momentum decay rate is a valid approximation that we can use in the Boltzmann equations. Moreover, equation \eqref{eq:DifferentialDecayRate} reduces to the correct Minkowskian decay rate when $n=0$.
\subsection{Boltzmann equations}
Having reviewed the necessary ingredients, we can now present and solve the Boltzmann equations governing the evolution of the energy densities.
We begin with the energy density $\rho_\phi$ of the massive $\phi$ particles. The integrated Boltzmann equation for this reads as \citep{Kolb_Turner}
	\begin{align}\label{eq:EnDensityPhi}
	\dot{\rho}_\phi(t)+3H(t)\rho_\phi(t)=-\Gamma_\chi(t)\rho_\phi(t)+w_\phi(t),
	\end{align}
where $H(t)=\dot{a}(t)/a(t)$ is the Hubble parameter. The second term on the left-hand side accounts for the dilution of the (non-relativistic) particles due to the expansion of the spacetime while the first term on the right-hand side accounts for their decay into massless $\chi$-particles. The last term $w_\phi$ on the right-hand side is the energy of the gravitationally created particles. Here, it is assumed that particle creation occurs only during the stiff matter dominated period and after the transition it is zero. The contribution of particle creation to energy density can be obtained from the differential creation rate  per mode $k$, equation \eqref{eq:DiffCreation},  by integration over all momenta as
	\begin{align}\label{eq:Gamma_Phi_integral}
	w_\phi(\eta)=\frac{1}{2\pi}\int_0^\infty dk k^2\frac{d|\beta_k|^2}{d\eta}\omega_k(\eta),
	\end{align}
where $\omega_k(\eta)=\sqrt{k^2+m^2b^2\eta}$. To perform the integration exactly, we make the approximation $\omega_k(\eta)\approx mb\eta^{1/2}$. The validity of this approximation was shown to be correct in \citep{Lankinen_Vilja:2017a} where it was noted that in a stiff matter dominated era the particle production rate is dominated by non-relativistic modes. With this approximation we can perform the integral in equation  \eqref{eq:Gamma_Phi_integral} exactly yielding
	\begin{align}
	w_\phi(\eta)=\frac{(mb)^{13/3}}{16}\eta^{3/2}\big[\Ai(-(mb)^{2/3}\eta)^2 +\Bi(-(mb)^{2/3}\eta)^2 \big].
	\end{align}
Switching back to standard coordinate time using the relation $a(\eta)d\eta=dt$, we obtain
	\begin{align}\label{eq:Gamma_phi}
	w_\phi(t)=\frac{3(mb)^{13/3}}{32b}t\big[\Ai(-(3mt/2)^{2/3})^2 +\Bi(-(3mt/2)^{2/3})^2].
	\end{align}		
For the differential equation \eqref{eq:EnDensityPhi} for $\rho_\phi$, we obtain a formal solution
\begin{align}\label{eq:rhophi}
	\rho_\phi(t)=\frac{1}{a(t)^3}e^{-\int_{t_0}^t \Gamma_\chi(t')dt'}\int_{t_0}^t a(t')^3w_\phi(t')e^{\int_{t_0}^{t'} \Gamma_\chi(t'')dt''}dt',
	\end{align}
where $t_0$ denotes the initial time taken to be the time when inflation ends. It has been assumed that the initial energy density $\rho_{\phi}(t_0)$ is zero as was discussed in the introduction.	
The integration in the exponential can be performed exactly yielding the decay rate in terms of Bessel functions,
	\begin{align}\nonumber\label{eq:DecayIntegral}
	f(t_0,t,n):&=\int_{t_0}^t \Gamma_\chi(t')dt'\\\nonumber\
	&=\frac{\lambda^2 t^2}{64}[J_\alpha(mt)^2-J_{\alpha-1}(mt)J_{\alpha+1}(mt)-Y_{\alpha+1}(mt)Y_{\alpha-1}(mt)+Y_{\alpha}(mt)^2]\\
	&-\frac{\lambda^2 t_0^2}{64}[J_\alpha(mt_0)^2-J_{\alpha-1}(mt_0)J_{\alpha+1}(mt_0)-Y_{\alpha+1}(mt_0)Y_{\alpha-1}(mt_0)+Y_{\alpha}(mt_0)^2],
	\end{align}		
where we have defined a function $f(t_0,t,n)$ as the integral of the decay rate from initial time $t_0$ to the time $t$ to simplify subsequent notation.
The evolution of the energy density $\rho_\chi$ of the massless $\chi$ particles is given by
\begin{align}
\dot{\rho_\chi}(t)+4H(t)\rho_\chi(t)=\Gamma_{\chi}(t)\rho_\phi(t),
\end{align}
where the second term on the left side represents dilution, in this case for relativistic particles, and the right hand side represents the creation of the $\chi$ particles as the massive $\phi$ particles decay into them. The formal solution is given as
\begin{align}\label{eq:rhopsi}
	\rho_\chi(t)=\frac{1}{a(t)^4}\int_{t_0}^t \Gamma_\chi(t')\rho_\phi(t')a(t')^4 dt'.
\end{align}
Because massless conformally coupled particles are not created by the expansion \citep{Parker:1969}, they are produced only as decay products. Since the initial energy density of the massive particles is zero, we have also set  $\rho_\chi(t_0)=0$ reflecting the fact that no decay has yet occurred.
Finally we note that the energy density of the background in the stiff matter era is given by \citep{Chavanis:2015}
\begin{align}\label{eq:rhostiff}
	\rho_{\mathrm{stiff}}(t)=\frac{1}{24\pi G_N t^2},
\end{align}
where $G_N$ is the gravitational constant.

\section{Reheating via gravitational particle production}\label{sec:3}
In this section we will describe the procedure to obtain the reheating temperature of the Universe and in the next section present the numerical results obtained from using this method.
Before proceeding further, we note that there are two different scenarios that may occur. Consider first the energy density $\rho_\phi$ of the gravitationally created particles. If the energy density of the massless particles $\rho_\chi$ grows slowly, the energy density of the massive particles reaches equilibrium with the background energy density $\rho_{\rm{stiff}}$ and the Universe ends up being temporarily matter dominated. This phase lasts until the energy density $\rho_\chi$ of the massless particles becomes dominant and the Universe transitions into radiation dominated one.

On the other hand, it may be that the energy density $\rho_\chi$ becomes dominant to matter energy density $\rho_\phi$ before they reach the energy density of the background $\rho_{\rm{stiff}}$. In this case, the Universe transitions into radiation dominated one when $\rho_\chi=\rho_{\rm{stiff}}$ skipping the temporary matter dominated era. Therefore, we must consider these two cases separately.

\subsection{Matter dominated era}
The Universe ends up as matter dominated if the energy density of matter is equal to the energy density of the background, provided that the energy density of the massive particles is greater than that of the massless particles (radiation) at that point. So, we start with the condition
	\begin{align}
	\rho_\phi=\rho_{\mathrm{stiff}}, \ \text{with the restriction} \ \rho_\chi<\rho_\phi.
	\end{align}
From this equality we obtain the time $t_{eq}$ at which the energy densities are equal and we can calculate the energy density at this time. The Universe becomes a matter dominated one and the evolution of the energy densities are described by ordinary Boltzmann equations without gravitational particle creation processes. The reason to neglect particle production in the matter dominated era is not only to make the model simpler, but mainly because to the knowledge of the authors the Bogoliubov coefficients for particle creation in matter dominated universe are not known. 

As the Universe expands in the matter dominated era, the energy density is given as a solution of ordinary Boltzmann equation:
	\begin{align}
	\rho_\phi^{mat}(t)=\Big(\frac{t_{eq}^{2/3}}{t^{2/3}}\Big)^3 e^{-f(t_{eq},t,4)}\times\rho_\phi(t_{eq}),
	\end{align}
where the superscript $mat$ indicates that this corresponds to the energy density in the matter dominated era.	 	
The function $f(t_0,t_{eq},n)$ now carries with it $n=4$ to indicate that the decay rate corresponds to a curved spacetime decay rate in matter dominated era. The massless particle energy density in the matter dominated era $\rho_\chi^{mat}$ is given by 
	\begin{align}\label{eq:RhoChiMat}
	\rho_\chi^{mat}(t)=\Big(\frac{1}{t^{2/3}}\Big)^4 \int_{t_{eq}}^t \Gamma_\chi(t')\rho_\phi^{mat}(t')(t'^{2/3})^4dt'+\rho_\chi(t_{eq})\Big(\frac{t^{2/3}_{\rm{eq}}}{t^{2/3}} \Big)^4,
	\end{align}
where the last term describes the dilution of the initial energy density calculated at the equilibrium.	
After this, the Universe continues to be matter dominated, until the energy densities of the massive and massless particles are equal. So once again, we equate the energy densities
	\begin{align}
	\rho_\chi^{mat}(t)=\rho_\phi^{mat}(t)
	\end{align}
to obtain some time $\tau_{eq}$, when they are equal. At this time the Universe transfers to radiation dominated era and we can calculate the energy density at this time, which is used as an initial condition. 
The energy densities in the radiation dominated phase are given by
	\begin{align}\label{eq:RhoChiRad}
	\rho_\phi^{rad}=\Big(\frac{\tau_{eq}^{1/2}}{t^{1/2}}\Big)^3 e^{-f(\tau_{eq},t,2)}\times\rho_\phi^{mat}(\tau_{eq})
	\end{align}
and
	\begin{align}
	\rho_\chi^{rad}(t)=\Big(\frac{1}{t^{1/2}}\Big)^4 \int_{\tau_{eq}}^t \Gamma_\chi(t')\rho_\phi^{rad}(t')(t'^{1/2})^4dt'+\rho_\chi^{mat}(\tau_{eq})\Big(\frac{\tau^{1/2}_{\rm{eq}}}{t^{1/2}} \Big)^4.
	\end{align}
The function $f$ now has $n=2$, which corresponds to radiation dominated era. The final step in the procedure is the calculation of the reheating temperature. To obtain this, we maximize the function $\rho^{rad}_\psi$ with respect to the time $t$. This gives us the reheating time $t_{rh}$ as well as the reheating temperature as $\frac{\pi^2}{30}g_*T_{rh}^4=\rho_{\chi,max}^{rad}$. In this equality, we neglect the constant terms and degrees of freedom $g_*$, as their numerical value is of order one when the fourth root is taken to obtain the temperature.

\subsection{Radiation dominated era}
The other situation arises if the energy density of the massless particles dominates that of massive particles as the energy densities reach equilibrium with the background energy density, i.e.,
	\begin{align}
	\rho_\chi=\rho_{\mathrm{stiff}}, \ \text{with the restriction} \ \rho_\chi>\rho_\phi.
	\end{align}
In this case the Universe ends up straight into radiation dominated universe. The massive particle energy density is given as a solution to the Boltzmann equations
	\begin{align}
	\rho_\phi^{rad}(t)=\Big(\frac{t_{eq}^{1/2}}{t^{1/2}}\Big)^3 e^{-f(t_0,t_{eq},2)}\times\rho_\phi(t_{eq}),
	\end{align}
with the index $n=2$ in the function $f$ indicating that we are in radiation dominated universe. The energy density for massless particles is given as
	\begin{align}
	\rho_\chi^{rad}(t)=\Big(\frac{1}{t^{1/2}}\Big)^4 \int_{t_{eq}}^t \Gamma_\chi(t')\rho_\phi^{rad}(t')(t'^{1/2})^4dt'+\rho_\chi(t_{eq})\Big(\frac{t^{1/2}_{\rm{eq}}}{t^{1/2}} \Big)^4,
	\end{align}
where $t_{eq}$ is the time when the background stiff matter energy density $\rho_{\mathrm{stiff}}$ and the energy density of the massless particles $\rho_\psi$ is equal.
The reheating time is obtained in the exactly same way by maximizing the radiation energy density.

\section{Numerical results}\label{sec:4}
Having established the procedure for obtaining the reheating temperature, we use Python programming to numerically evaluate the integrals of the preceding section. This requires us to fix several parameters of the model as well as to choose proper units. A natural choice would be to express the results using natural units i.e., in units where $\hbar=c=1$, but this presents us with a problem regarding the numerical simulation. In natural units, the numerical values range from very small to very large with tens of orders of magnitude in difference. This makes the numerical simulation prone to errors, increases computational time and prevents the calculation of full range of parameters. To remedy this situation, we will use Planck units where $G_N=1$ and the results can be expressed in natural units after calculations.

There are a total of five parameters which must be fixed prior to calculations. We must fix the initial time $t_0$, which marks the end of inflation,  the mass $m$ of the decaying particle, the parameter $b$ and the values of the coupling constants $\lambda$ and $\xi$. We will first discuss about the values where these parameters were set and after that discuss about the reheating temperature and some features which arise from the simulation.

\subsection{Fixing the parameters}
The initial time $t_0$ is fixed to be the time when inflation effectively ends. Although this time is not precisely known, we fixed the initial time at $t_0=10^{11}$ corresponding to $t_0\sim 10^{-32}\ \text{sec}$, a value found commonly used in literature \citep{Kolb_Turner,Liddle_Lyth}. We checked how this particular choice affects the results by running the simulation with different values varying a few order of magnitude around the value $t_0=10^{11}$. It was found that the change in initial time presents no observable changes in the numerical values of the reheating temperature although it does affect somewhat the transition time to matter and radiation phases. The results in the rest of this paper all have the initial value set to $t_0=10^{11}$. 

There are two other parameters which did not produce any observable effects on the reheating temperature when running the simulation. These are the gravitational coupling $\xi$ and the parameter $b$. Although not affecting on the reheating temperature itself, these parameters have effect on other features like the time of reheating and time of transition to different phases of the Universe. The parameter $b$ was calculated on the range of $b\in[10^{-1}, 10^{1}]$. For the gravitational coupling, we chose to use minimal and conformal coupling.  The simulation was also ran with other values of $\xi$ lying between the minimal and conformal coupling resulting in no observable changes in reheating temperature.

The mass was constrained to the widest range possible which did not create errors in the numerical integration and was found to be in the range from $10^{-17}$ to $10^{-9}$ in Planck units. This corresponds to a mass range of about $100\ \text{GeV}$ to $10^{10}\ \text{GeV}$, which provides a fairly large account of physically interesting masses. The coupling $\lambda$ is assumed to be small, in order for the perturbative expansion in $\lambda/m$ to work. It was fixed to run with the mass through the relation $\lambda=\gamma m$, with three values for $\gamma$, namely $\gamma=10^{-1}, 10^{-2}, 10^{-3}$. Moreover, the plots were produced with a grid of $200 \times 200$ points of the parameters $m$ and $b$.

\subsection{Reheating temperature}
The simulation was ran with the parameters fixed as described above. Before presenting the results for the reheating temperature, we want to emphasize an interesting feature which the simulation presented. As noted in section \ref{sec:3}, there exists two possibilities for the evolution of the Universe after the stiff matter era: it either ends up being matter dominated for a while or becomes immediately radiation dominated. With the full range of parameters used, we found that in every situation the Universe always ends up being temporarily matter dominated before the radiation dominated era. This implies that the decay into massless particles is not fast enough to increase their energy density sufficiently above matter energy density before equilibrium with the background is reached. A direct transition into radiation dominated universe might be possible by increasing the coupling $\lambda$ sufficiently. However, for a large value the perturbative expansion might not work anymore.

		\begin{figure}[t]
	\centering
	\mbox{\subfigure[]{\includegraphics[scale=0.5]{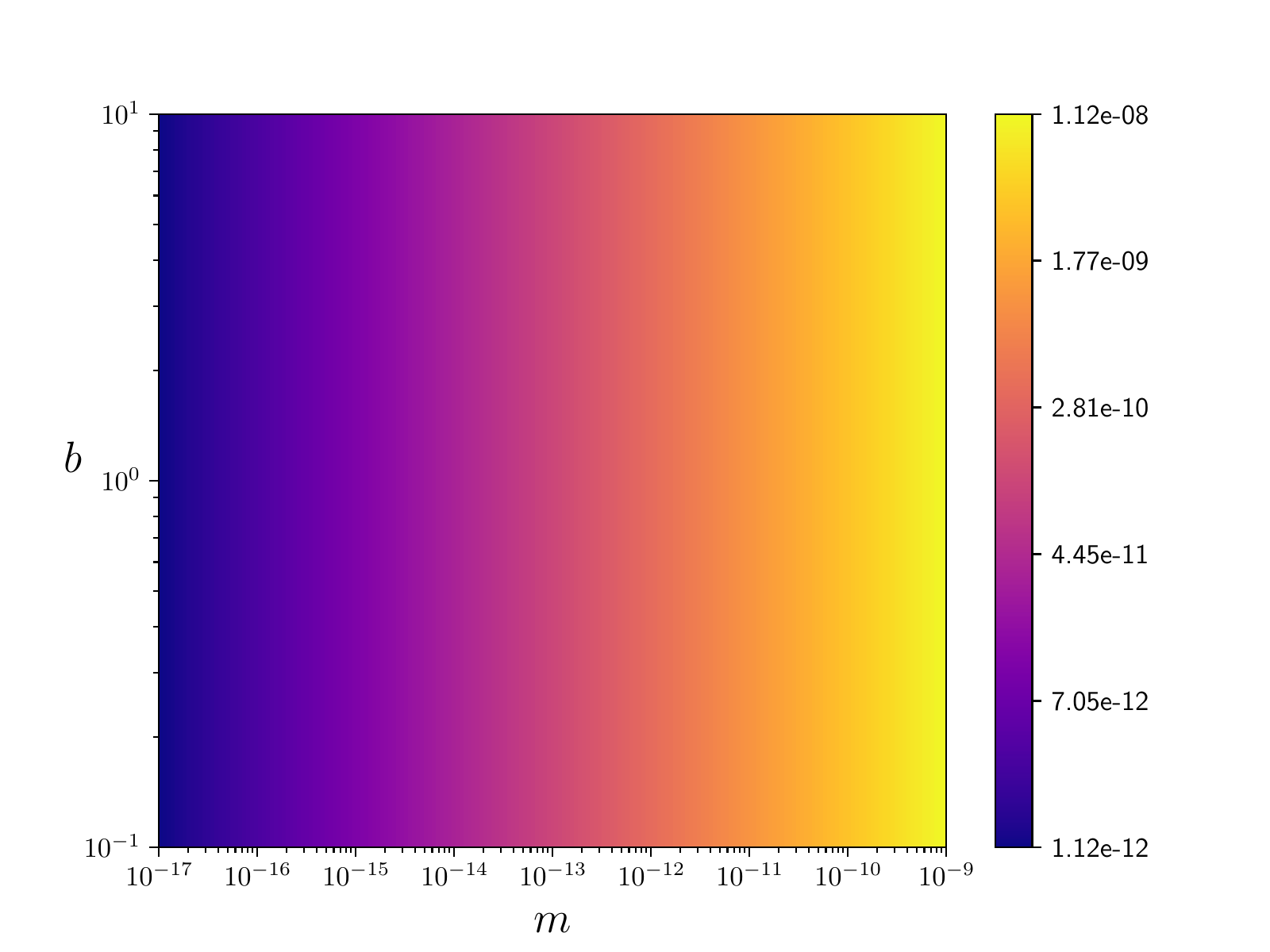}}\subfigure[]{\includegraphics[scale=0.5]{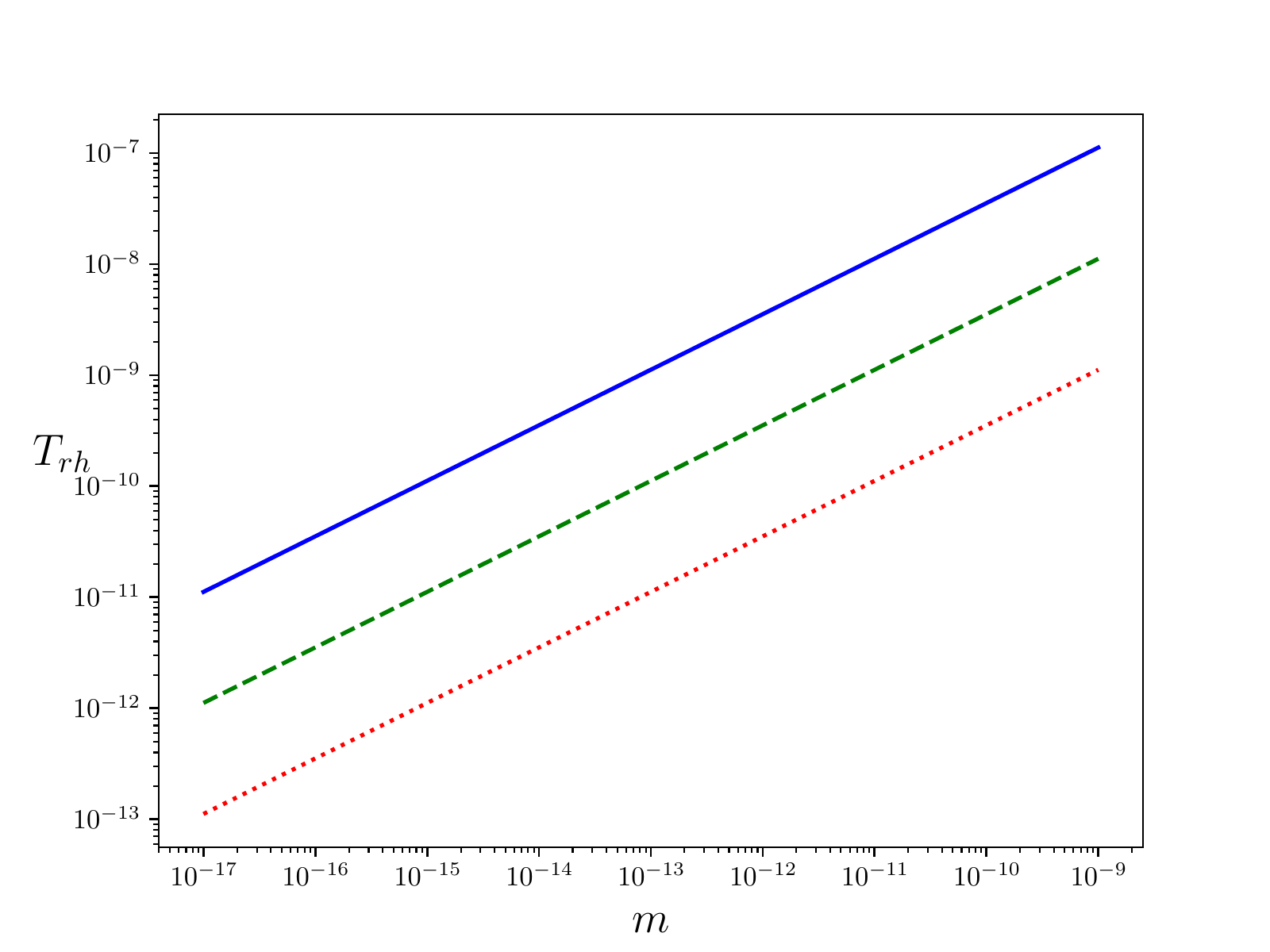}}}
	\caption{(a) Reheating temperature $T_{rh}$ as a function of mass $m$ and expansion parameter $b$ given in Planck units with the ratio $\lambda/m=10^{-2}$ and $\xi=1/6$ and (b) Reheating temperature in Planck units as a function of mass $m$ with three different values of the ratio $\lambda/m$, namely $10^{-1}$ (blue, solid), $10^{-2}$ (green, dashed) and $10^{-3}$ (red, dotted) with $b=10$ and $\xi=1/6$.}\label{fig:masstemp} 
	\end{figure} 
For the actual reheating temperature, we found it to be in the interval of about $10^{-13}-10^{-7}$ in Planck units depending on the values of the parameters used. This corresponds to about order of $10^6-10^{12}\ \text{GeV}$. Moreover, as was precedingly noted, the reheating temperature is practically independent of the coupling $\xi$ and the parameter $b$. The independence of the reheating temperature from the parameter $b$ in the matter dominated case is manifest in figure \ref{fig:masstemp}a, which plots the temperature as a function of the mass $m$ and the parameter $b$. 
The coupling parameter $\lambda$, or rather its ratio to mass, does have an effect on the reheating temperature. With fixed $m$, an increase in this ratio corresponds to increasing the coupling constant $\lambda$ by the same amount.  As we raise the coupling $\lambda$ by an order of magnitude, the reheating temperature is raised by an order of magnitude correspondingly (figure \ref{fig:masstemp}b). To understand why this is so, we can take a look at equation \eqref{eq:DifferentialDecayRate} from where it can be seen that the parameter $\lambda$ affects the decay rate of the particles. As $\lambda$ is increased the decay rate increases and there is faster decay into the massless particles. This in turn increases the energy density $\rho_\chi$ as seen from equations \eqref{eq:RhoChiMat} and \eqref{eq:RhoChiRad}.  Therefore, as the energy density of $\chi$ particles is higher to begin with when increasing the coupling $\lambda$, it is natural consequence that the reheating temperature is correspondingly higher.

\begin{figure}[t]
	\centering
	\mbox{\subfigure[]{\includegraphics[scale=0.5]{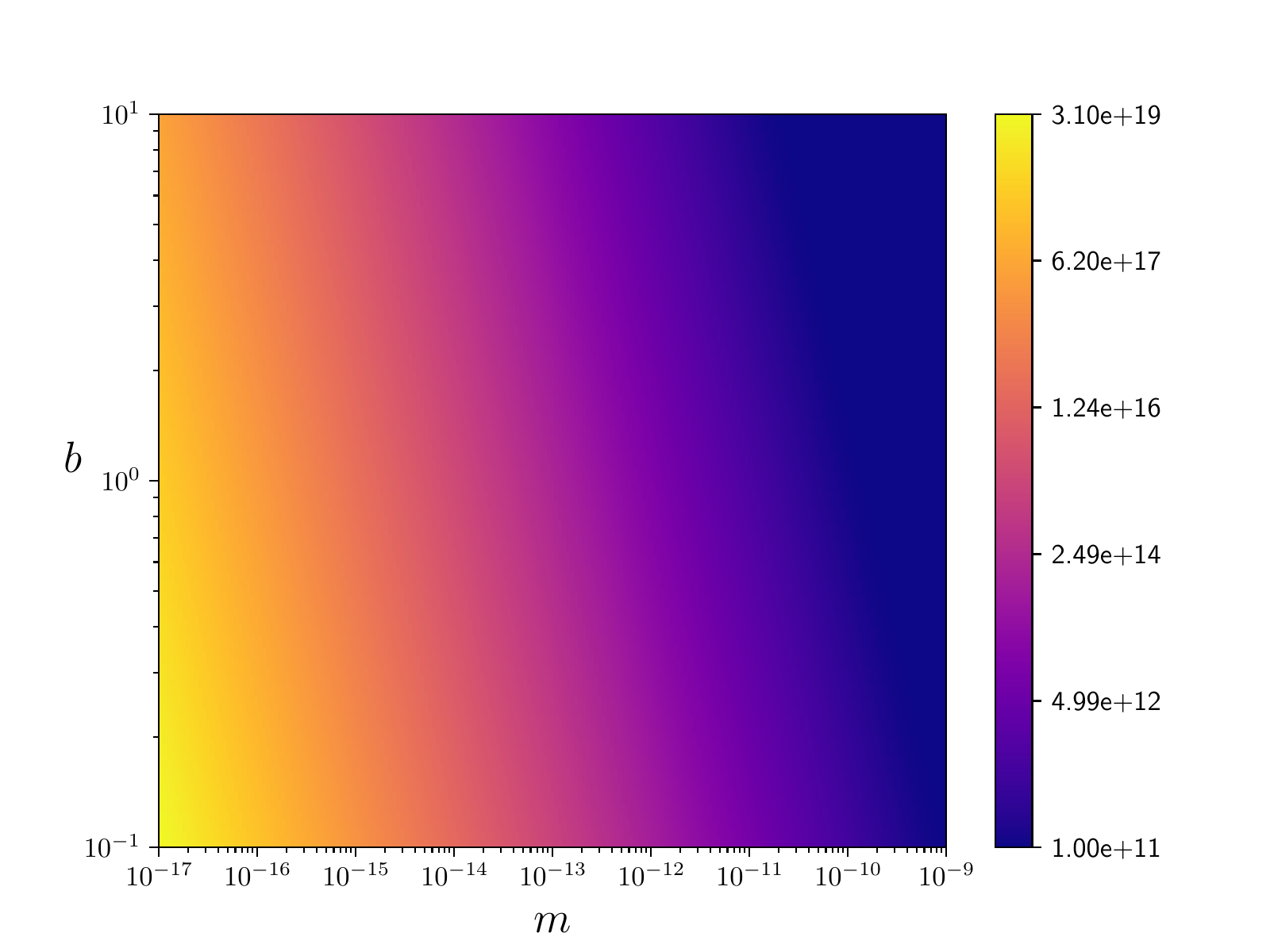}}\subfigure[]{\includegraphics[scale=0.5]{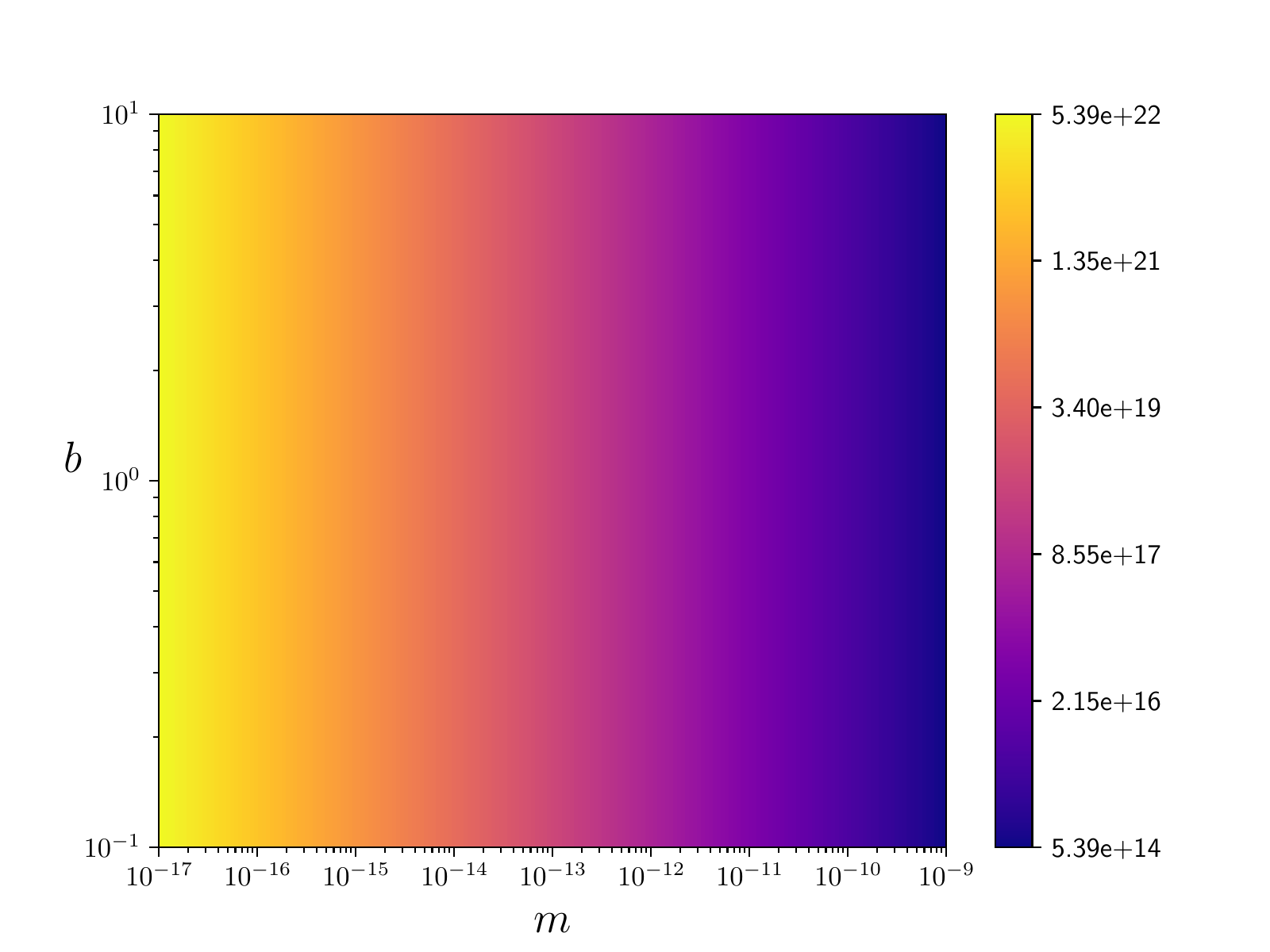}}}
	\caption{(a) Time of transition to matter dominated era as a function of mass $m$ and the expansion parameter $b$ given in Planck units. (b) Reheating time as a function of mass $m$ and the expansion parameter $b$ in Planck units. In both figures the ratio $\lambda/m$ is $10^{-2}$ and the coupling $\xi$ is conformal. }\label{fig:bparametri} 
	\end{figure}

The expansion rate $b$ does have an effect on the time when the Universe reaches matter dominated era and it has no effect on the transition time from matter to radiation dominated era. Figure \ref{fig:bparametri}a gives a clearer picture, where the time of transition to the matter dominated era is given as a function of the mass of the particle and the expansion rate $b$. As can be seen, increasing the expansion rate generally shortens the time when matter domination is reached. The same thing happens if the mass of the decaying particle increases. These are reasonable to expect because in \citep{Lankinen_Vilja:2017a} it was shown that in a stiff matter dominated universe particle creation is most effective for a very massive scalar field and large expansion parameter. Therefore, for a very massive particle, the particle creation is so explosive that the Universe reaches the temporary matter dominated era almost instantly.

Figure \ref{fig:bparametri}b shows the reheating time calculated as the time when the maximum energy density of the massless particles is reached. This is found to coincide with the time when the transition into radiation domination occurs. The maximum energy density therefore occurs at the transition and afterwards it begins to decrease due to dilution. The reheating time is found to be about two to three orders of magnitude larger than the corresponding time of transition to the matter dominated era. This time interval is the amount of time the Universe spends in the matter dominated era. What is different from the transition time to matter dominance is that the reheating time shows no dependence on the expansion parameter $b$. Some insight for this behavior can be gained by taking also into account the decay rate of the massive scalar. First of all, the parameter $b$ affects only the evolution of the energy densities in the stiff matter phase. From previous works on particle decay, we know that the decay rate for the massive scalar to decay into massless scalars is faster the more massive the particle is \citep{Lankinen_Vilja:2017b,Lankinen_Vilja:2018b}. It is therefore reasonable to expect that both energy densities, massive and massless, follow closely one another in the stiff matter phase regardless of the mass of the particle. After the transition to matter phase, the energy densities evolve independent of the parameter $b$, which would explain the independence of reheating time from this parameter in figure \ref{fig:bparametri}b.

\section{Discussion and outlook}\label{sec:5}
In this paper, we have presented a novel approach using Boltzmann equations and results from quantum field theory in curved spacetime. We have made some assumptions and approximations in our model which we wish to discuss further along with some interesting features arising from the model.

First, concerning the gravitational particle creation, we have assumed that it is abruptely stopped after the Universe changes phase from the stiff matter dominated one. This assumption amounts to a sudden transition into radiation and matter eras. It is, however, possible that there is some stiff matter left after the equilibrium time. This leftover stiff matter scales as $\rho_{\rm{stiff}}\propto a^{-6}$ so it is quickly diluted away. Therefore we do not consider the particle creation from the residual stiff matter to be significant enough to have an effect on the results. Also the use of Bogoliubov coefficients neglects the back reaction effect. The reaction of the particle creation back on the gravitational field would modify the expansion in such a way as to reduce the creation rate \citep{Parker:1969}. This effect would probably lower the reheating temperature somewhat because there are less particles overall to decay. A full treatment of this issue is beyond the scope of this paper.

Secondly, we have used the method of added-up probabilities which method gives a way to calculate decay rate in curved space which closest resembles the decay rate as defined in Minkowskian space. This method is somewhat restrictive because it requires conformally coupled massless particles, which are not gravitationally created in conformally flat metric, in order for the decay rate to make sense \citep{Audretsch_Spangehl:1985}. If there is in addition creation of massless non-conformally coupled particles from the vacuum, this would have effect on the energy density of the massless particles. What the magnitude of this effect to the numerical calculations would be we cannot say, but it does present an interesting point for future studies.

We have also found that the reheating temperature does not depend on the expansion parameter $b$ and the coupling parameter $\xi$ although these have an effect on when the Universe changes to the temporary matter dominated era. Nevertheless, these parameters most probably have an effect on cosmological aspects of the Universe, e.g., cosmological perturbations. This is an aspect we have not considered on our approach and model, but may be considered further in future research.

We can also take a look at some differences in our model and previous studies regarding reheating in the kination regime. In our approach, we used particle creation during the stiff matter dominated era in contrast to previous studies where the particle creation has been obtained by a change of metric. This difference in the models does not therefore allow a direct comparison of results, because in the previous studies the reheating temperature usually depends on the inflation scale and the transition time from inflation to kination \citep{Chun:2009}. Our model is independent of such parameters and offers a numerical value to the reheating temperature. In some previous studies, where particles were created by a sudden change of metric, the authors found the reheating temperature to be much smaller, just few orders of magnitude in $\text{GeV}$ \citep{Salo:2017a,Salo:2017b}. In our model, where we used gravitational particle creation during the expansion of spacetime, a much higher temperature is reached. This would imply that reheating via gravitational particle creation is more effective than the situation where matter is created solely from the sudden change of metric. We must advice on some caution on making this assumption, however, because the methods of calculation are different.

Finally, we wish to compare the differences of using curved spacetime decay rates as opposed to Minkowskian decay rates. As was noted, the differential decay rate \eqref{eq:DifferentialDecayRate} reduces to the Minkowskian decay rate in the correct limit $n=0$. We ran the simulation also with the Minkowskian decay rate allowing a comparison with the two approaches. It was found that Minkowskian and curved space decay rates produce results within the same order of magnitude with only small numerical differences. Although it is known that decay rates in curved space are modified from their Minkowskian counterparts \citep{Lankinen_Vilja:2018a,Lankinen_Vilja:2018b,Boyanovsky}, it is likely that the timescales in question are so small that the effect of curved space modification has not produced noticeable effects yet.

In summary, we have in this paper aimed for calculating reheating temperature of the Universe during kination regime via gravitational particle creation during expansion of the Universe. We used curved space decay rates for the scalar field to decay into radiation and found the reheating temperature to be in the range of $10^6-10^{12}\ \text{GeV}$. This temperature turns out to be independent on the expansion rate of the Universe and the coupling of the decaying particle into gravity. In this paper we considered a decay into scalar channel, but it is also possible to use a fermionic decay channel, for which the decay rates in curved spacetime were recently obtained \citep{Lankinen_Malmi:2019}. These two decay channels could be combined to study reheating more in detail. We leave this aspect also to future studies.

\begin{acknowledgments}
J.L. would like to acknowledge the financial support from the University of Turku Graduate School (UTUGS). O.K. would like to acknowledge the financial support from the Turku University Foundation.
\end{acknowledgments}

\end{document}